\documentclass[lettersize,journal]{IEEEtran}

\usepackage{colortbl}
\usepackage{supertabular}
\usepackage{cite}
\usepackage{xurl}
\usepackage{hyperref}

\usepackage{listings}
\usepackage{multirow}
\usepackage{algorithm}
\usepackage{algorithmicx}
\usepackage{algpseudocode}
\usepackage{graphicx}
\usepackage{subcaption}

\lstdefinelanguage{Solidity}{
	keywords=[1]{anonymous, assembly, assert, balance, break, call, callcode, case, catch, class, constant, continue, constructor, contract, debugger, default, delegatecall, delete, do, else, emit, event, experimental, export, external, false, finally, for, function, gas, if, implements, import, in, indexed, instanceof, interface, internal, is, length, library, log0, log1, log2, log3, log4, memory, modifier, new, payable, pragma, private, protected, public, pure, push, require, return, returns, revert, selfdestruct, send, solidity, storage, struct, suicide, super, switch, then, this, throw, transfer, true, try, typeof, using, value, view, while, with, addmod, ecrecover, keccak256, mulmod, ripemd160, sha256, sha3}, 
	keywordstyle=[1]\color{blue}\bfseries,
	keywords=[2]{address, bool, byte, bytes, bytes1, bytes2, bytes3, bytes4, bytes5, bytes6, bytes7, bytes8, bytes9, bytes10, bytes11, bytes12, bytes13, bytes14, bytes15, bytes16, bytes17, bytes18, bytes19, bytes20, bytes21, bytes22, bytes23, bytes24, bytes25, bytes26, bytes27, bytes28, bytes29, bytes30, bytes31, bytes32, enum, int, int8, int16, int24, int32, int40, int48, int56, int64, int72, int80, int88, int96, int104, int112, int120, int128, int136, int144, int152, int160, int168, int176, int184, int192, int200, int208, int216, int224, int232, int240, int248, int256, mapping, string, uint, uint8, uint16, uint24, uint32, uint40, uint48, uint56, uint64, uint72, uint80, uint88, uint96, uint104, uint112, uint120, uint128, uint136, uint144, uint152, uint160, uint168, uint176, uint184, uint192, uint200, uint208, uint216, uint224, uint232, uint240, uint248, uint256, var, void, ether, finney, szabo, wei, days, hours, minutes, seconds, weeks, years},	
	keywordstyle=[2]\color{teal}\bfseries,
	keywords=[3]{block, blockhash, coinbase, difficulty, gaslimit, number, timestamp, msg, data, gas, sender, sig, value, now, tx, gasprice, origin},	
	keywordstyle=[3]\color{violet}\bfseries,
	identifierstyle=\color{black},
	sensitive=true,
	comment=[l]{//},
	morecomment=[s]{/*}{*/},
	commentstyle=\color{gray}\ttfamily,
	morestring=[b]',
	morestring=[b]",
  moredelim=[is][\color{red}]{@@}{@@},
  moredelim=[is][\color{green}]{!!}{!!},
  commentstyle=\color{purple}\ttfamily,
}

\usepackage{lipsum}
\usepackage{pythonhighlight}

\usepackage{amsmath,amsopn,amssymb}
\usepackage{endnotes,microtype,xspace,graphicx,fancyvrb,multirow}
\usepackage{booktabs}
\usepackage{array,relsize}
\usepackage[T1]{fontenc}
\usepackage{times}
\usepackage{fancyhdr}

\usepackage[labelfont=bf,font=small,skip=5pt]{caption}
\usepackage{fp}
\usepackage{siunitx}

\usepackage{minted}

\usepackage{balance}

\usepackage{subcaption} 
\usepackage{latexsym}

\usepackage{amsthm}
\usepackage{mdframed}

\newtheorem{definition}{Definition} 

\newtheorem{theorem}{Theorem}

\newtheorem{proposition}{Proposition}

\usepackage{minted}
\setminted{fontsize=\small, autogobble=true, breaklines}
\usepackage{tikz}
\newcommand*{\circled}[1]{\lower.7ex\hbox{\tikz\draw (0pt, 0pt)%
    circle (.5em) node {\makebox[1em][c]{\small #1}};}}

\usepackage{markdown}
\usepackage{bbding}

\markdownSetup{fencedCode = true}

\newcommand{\sys}{\textsc{FAuditor}\xspace}

\fvset{fontsize=\scriptsize,xleftmargin=8pt,numbers=left,numbersep=5pt}

\input{code/fmt}

\def\Snospace~{\S{}}

\algrenewcommand{\algorithmiccomment}[1]{\hfill{\(\triangleright\) #1}}

\newif\ifdraft\drafttrue
\newif\ifnotes\notestrue
\ifdraft\else\notesfalse\fi

\input{glyphtounicode}
\newcolumntype{R}[1]{>{\raggedleft\let\newline\\\arraybackslash\hspace{0pt}}p{#1}}

\newcommand{\squishlist}{
\begin{itemize}[noitemsep,nolistsep]
  \setlength{\itemsep}{-0pt}
}
\newcommand{\squishend}{
  \end{itemize}
}

\usepackage{tikz}

\usepackage{xstring}
\newcommand{\PP}[1]{
\noindent{\bf \IfEndWith{#1}{.}{#1}{#1.}}
}

\newcommand{\boxbeg}{
\vspace{2px}
\noindent\begin{tabular}{|l|}\hline
\begin{minipage}{3.2in}
\vspace{2px}
\noindent
}

\newcommand{\boxend}{
\vspace{2px}
\end{minipage}\\ \hline
\end{tabular}
\vspace{-10pt}
}

\usepackage{soul} 
\usepackage[strings]{underscore}

\begin{document}
\title{Capturing Monetarily Exploitable Vulnerability in Smart Contracts via Auditor Knowledge-Learning Fuzzing}



\ifdefined\DRAFT
 \pagestyle{plain}
 \lhead{Rev.~\therev}
 \rhead{\thedate}
 \cfoot{\thepage\ of \pageref{LastPage}}
\fi


\author{
    \IEEEauthorblockN{Bowen Cai$^1$, Weiheng Bai$^1$, Hangyun Tang$^2$, Youshui Lu$^3$, Kangjie Lu$^1$}\\
    \IEEEauthorblockA{$^1$University of Minnesota - Twin Cities, $^2$Fudan University, $^3$Xi'an Jiaotong University \\ 
    \{cai000254, bai00093\}@umn.edu, \{23110240090\}@m.fudan.edu.cn,\\\{yolu6176\}@uni.sydney.edu.au, \{kjlu\}@umn.edu}
}





\ifdefined\DRAFT
 \pagestyle{plain}
 \lhead{Rev.~\therev}
 \rhead{\thedate}
 \cfoot{\thepage\ of \pageref{LastPage}}
\fi


\newcommand{\UMNaffil}{%
  \affiliation{%
    \institution{University of Minnesota}
    \city{Minneapolis}
    \state{MN}
    \country{USA}
  }%
}

\newcommand{\XJTUaffil}{%
  \affiliation{%
    \institution{Xi'an Jiaotong University}
    \city{Xi'an}
    \state{Shaanxi}
    \country{China}
  }%
}

\newcommand{\JHUaffil}{%
  \affiliation{%
    \institution{Johns Hopkins University}
    \city{Baltimore}
    \state{Maryland}
    \country{USA}
  }%
}

\newcommand{\ZJUaffil}{%
  \affiliation{%
    \institution{Zhejiang University}
    \city{Hangzhou}
    \state{Zhejiang}
    \country{China}
  }%
}

\date{}
 
\maketitle
\sloppy

\begin{abstract}
Smart contracts have extended blockchain functionality beyond simple transactions, powering complex applications like decentralized finance (DeFi). However, this complexity introduces serious security challenges, including price manipulation and inflation attacks. Despite the development of various security tools, the rapid rise in financially motivated exploits continues to pose a significant threat to the blockchain ecosystem.

These financially motivated exploits often stem from Monetarily Exploitable Vulnerabilities (MEVuls), which refer to vulnerabilities arising from exploitable implementations in monetary transactions or value-transfer logic. Due to their complexity, intricate chains of function calls, multifaceted logic, and diverse manifestations across different smart contracts, MEVuls are particularly challenging for current security tools to identify. Instead of providing actionable insights, existing tools frequently generate excessive warnings that overwhelm developers without effectively mitigating risks.

To address the challenge of recognizing MEVuls, we first formalize MEVuls based on common real-world financial exploits. Then, we introduce \textsc{FAuditor}, a specialized fuzzer designed to detect MEVuls in smart contracts. The key insight is that leveraging smart contracts' finance-related interfaces directly exposes critical vulnerabilities, making detection more targeted. \textsc{FAuditor} constructs an oracle for four core MEVul types by focusing on these interfaces. We further integrate auditors' reports using NLP to extract valuable insights on exploitation patterns, enabling a more informed search strategy. Additionally, \textsc{FAuditor} employs a self-learning mechanism that refines its detection strategies over time, allowing it to improve based on prior fuzzing results.
In our evaluation, \textsc{FAuditor} impressively reveals 220 zero-day MEVuls. Meanwhile, compared to existing fuzzers, \textsc{FAuditor} detects vulnerabilities faster and achieves better instruction coverage.
\end{abstract}

\section{Introduction}

The blockchain technology, as a decentralized and distributed ledger, underpins digital currencies like Bitcoin and Ethereum. A key feature of Ethereum is the smart contract~\cite{noauthor_ethereum_nodate}, a self-executing program that extends blockchain functionality and is widely used in applications such as decentralized exchanges~\cite{xu_sok_2023, immunefi_balancer_2023}, lending protocols~\cite{zhou_just--time_2021, li_survey_2022}, and digital insurance~\cite{chen2021traceable,xu2020improving}. However, their association with monetary transactions and public accessibility exposes smart contracts to significant security risks, including price manipulation and inflation attacks, which resulted in losses exceeding \$5.4 billion from 2022 to 2023~\cite{team_stolen_2024}. As a result, addressing smart contract vulnerabilities is a critical priority for both industry and academia.

Automated fuzzing tools have significantly advanced smart contract security by mitigating traditional vulnerabilities such as reentrancy, under/overflow errors, and unsafe delegate calls~\cite{luu_making_2016, Groce2021a, Feist2019, wesley_verifying_2022}. Early fuzzing techniques used random input generators, while recent research has introduced path constraints to explore deeper code layers~\cite{Choi2021,liu_rethinking_2023}. Additionally, machine and deep learning models have improved fuzzing efficiency, as seen in tools like ILF~\cite{he_learning_2019} and Skyfire~\cite{wang_skyfire_2017}. Fuzzing remains essential in smart contract development, providing real-time scenarios for detecting security vulnerabilities.

\PP{Unawareness of Monetary Exploitability}
Monetarily exploitable vulnerabilities (MEVuls) in smart contracts constitute flaws within the contract code that attackers can exploit to deviate the contract's intended functionality, leading to severe financial losses~\cite{atzei2017survey}. 
Despite the proliferation of security tools, significant financial losses due to smart contract vulnerabilities continue unabated, as highlighted in recent studies~\cite{bartoletti_solvent_2024, li_static_2024, yang_uncover_2024}. 
Perez~\cite{Perez2021a} highlights substantial discrepancies between the capabilities of existing security tools and their effectiveness in addressing MEVuls.

The MEVuls-unware issue manifests itself in two aspects. First, the proportion of exploitable vulnerabilities reported by existing tools is remarkably low; Perez~\cite{Perez2021a} noted that merely 1.98\% of reported vulnerabilities are actually exploitable. This trend persists across various types of vulnerabilities. Second, a significant portion of MEVuls remains undetected by these tools. As noted by Chaliasos~\cite{stefanov_towards_nodate}, of the 127 identified MEVuls, only 25\% were detected by current tools. As a result, current tools' reports are often filled with ambiguous information, leading developers to waste time identifying non-critical details and overlooking real dangers.

The persistence of the MEVuls-unaware issue arises from two primary causes: (1) \textit{misjudging the origin of exploitations} and (2) \textit{inadequacy of detection capability}. 
First, as indicated in existing studies~\cite{bartoletti_solvent_2024,stefanov_towards_nodate,immunefi_website,noauthor_immunefi_nodate}, MEVuls predominantly originate from high-level business logic errors, rather than the low-level opcode vulnerabilities targeted by existing tool. 
For instance, while most fuzzing tools focus predominantly on opcode and stack analysis~\cite{Choi2021,liu_rethinking_2023}, employing pattern matching to identify opcode vulnerabilities~\cite{godefroid_learnfuzz_2017,su_effectively_2022}, such patterns are not necessarily indicative of monetary exploitability~\cite{bartoletti_solvent_2024, li_static_2024, yang_uncover_2024}. 
Second, the detection of vulnerabilities is further complicated by their increasing complexity and sophistication, which often require intricate, multifaceted transaction sequences that surpass the capabilities of conventional tools~\cite{bartoletti_solvent_2024}.
For example, in \autoref{fig:article}, a human auditing report by Immunefi~\cite{immunefi_website} illustrated a proof of concept (PoC) involving over five transactions to reproduce the exploitation, a scenario that surpasses the detection threshold of tools like Mythril~\cite{ma_pluto_2022}, which typically times out after search depth of three transactions~\cite{immunefi_website}.

In this paper, we introduce \sys, a MEVul-aware fuzzer designed to address the common issue of tools being unaware of MEVuls. Our tool leverages key variables extracted from smart contract Application Binary Interfaces (ABIs), as these variables often hold critical information that may lead to financial exploits. \sys uses rules derived from auditors' reports, which capture real-world exploitation patterns, to generate function sequences that are likely to trigger MEVuls. The process has three key stages: (1) generating MEVul-aware functions based on auditor-derived rules to focus on high-risk areas, (2) detecting MEVuls by analyzing critical variables, which are typically central to monetary exploits, and (3) refining the rules' applicability by incorporating feedback from detected vulnerabilities, ensuring continuous improvement of the fuzzing strategy.

\PP{Challenges}
We identify two key challenges in implementing \sys. (1) \textit{Defining and Detecting MEVuls (C1).} MEVuls lack a standardized definition, making it difficult to establish consistent detection criteria. This variability complicates the creation of a robust detection system. \sys addresses this by formulating precise criteria that capture the financial exploitation potential in smart contracts. (2) \textit{Search Strategy for MEVuls (C2).} MEVuls often involve long, complex call chains that general-purpose fuzzers struggle to identify. Unlike simpler vulnerabilities, MEVuls require targeted strategies to explore deeper contract interactions. To tackle this, \sys incorporates a specialized search strategy focused on efficiently generating function sequences that are tailored to uncover these vulnerabilities within a practical timeframe.

\PP{Techniques}
To address C1, we introduce the first \textit{Formalization of MEVul and an Automatic Detector}. MEVuls are challenging to detect due to their lack of a standard definition, so we formalize them into four common types of financially damaging exploits observed in real-world scenarios. This allows us to develop a generalized detector that can automatically apply consistent detection logic across various contracts, ensuring reliable identification of these vulnerabilities.
For C2, we propose a \textit{Rule-Based and Self-Learning Strategy for Function Generation}. MEVuls often require precise function sequences to expose, which are not easily captured by general fuzzing strategies. Our approach leverages a dataset compiled from auditors' reports to generate function sequences targeting high-risk, MEVul-related interactions. The strategy is further refined using a self-learning mechanism that adapts based on the outcomes of previous fuzzing attempts, improving its effectiveness over time.

We extensively evaluated \sys by compiling a dataset of 441 bug review articles sourced from an auditing website~\cite{immunefi_website, noauthor_code4rena_nodate, noauthor_analysis_nodate}, along with the corresponding smart contract codebases. 
This dataset was utilized to train our NLP model and develop the Rule-Based Strategy, Self-Learning Strategy, and MEVul detection module. 
During the evaluation, we tested our fuzzer on a dataset comprising 15,132 individual smart contracts and 23 decentralized applications (DApps) from Ethereum. 
Consequently, our fuzzer identified \textbf{213} new vulnerabilities in individual smart contracts and \textbf{7} new vulnerabilities in DApps, which were not detected by existing tools. 
Our fuzzing method demonstrates a 27\% improvement in coverage efficiency on DApps, compared to state-of-the-art (SOTA) methods, and it also excels in faster bug discovery, particularly for 6 out of 9 vulnerability types (including 5 traditional types and 4 MEVuls).

\PP{Contributions} In summary, our contributions are:

\noindent $\bullet$ \textbf{MEVul Formalization.} We introduce the first formalization of MEVuls to fill the gap between current research and Real-world Exploitable Vulnerability (MEVul). By summarizing critical exploitation results and understanding attackers' motivations, we categorize MEVuls into four types, encompassing the most financial losses-related scenarios in smart contracts.

\noindent $\bullet$ \textbf{MEVul-aware Fuzzing Tool.} We propose a fuzzing tool to tackle the challenge of being unaware of the MEVuls. This is achieved through the innovation of (i) new casting oracles that capture MEVuls, (ii) a function sequence generation strategy that leverages auditors' reports to improve the efficiency of MEVul detection, and (iii) a new feedback mechanism that optimizes the function generation based on detection history.

\noindent $\bullet$ \textbf{Prototype and Evaluation.} We developed and validated our prototype system using a dataset of 15,132 individual smart contracts and 23 new Dapps on Ethereum. The validation revealed 213 vulnerabilities in individual contracts and 7 vulnerabilities in Dapps, demonstrating the system's outstanding performance in detecting exploitable vulnerabilities and higher efficiency in code coverage than SOTA tools. Our tool will be open-sourced upon publication.

\section{Background}


\subsection{Smart Contract Fuzzing}\label{sec:back_tradition}

The fuzzing method is an automated testing technique that involves providing random inputs to a program in order to generate real test cases for smart contracts. It has become an essential testing phase in smart contract development for preventing real-world threats. Recent work has primarily focused on progressing in two key areas: (i) the search strategy and (ii) the vulnerability oracle.


\PP{Search Strategies of Fuzzing Tools}
The search strategy refers to the methods used by fuzzing tools to generate and select inputs. Most traditional smart contract fuzzers use the random strategy~\cite{ali_sescon_2021}, which is neither efficient nor accurate. Recently, research projects like SmarterContract, Smartian, etc.~\cite{sendner_smarter_2023, Choi2021, godefroid_learnfuzz_2017, jiang_contractfuzzer_2018, liu_rethinking_2023, zhang_deltafuzz_2022} start incorporating the machine learning model, remarkably improving fuzzing tools' feasibility. Their common practice is to sample the results from static analysis and then use the machine learning model to direct the input selection. The reasoning behind them is to imitate the static analyzer's optimum execution path and only test the most risky branches of the code, improving the efficacy a lot. Meanwhile, some other fuzzers incorporate advanced models like reinforcement learning~\cite{abdelaziz_smart_2023} and convolutional network~\cite{he_learning_2019} to help the fuzzer select inputs. This work did help search for deeper layer vulnerabilities and improve efficacy. However, they struggle to generate input that can trigger MEVuls because they lack recognition of MEVuls and are missing related datasets.


\PP{Traditional Vulnerability Oracles}
Oracle is the module that identifies vulnerability types in the smart contract code. Traditionally, the focus of oracles mainly refers to the vulnerabilities recorded by SWC\cite{swc_website}, an official vulnerability repository maintained by the Ethereum Foundation. For example, it includes classical vulnerabilities like Unhandled Exceptions (UE), Re-entrance (RE), Suicidal (SC), Block State Dependency (BD), Leaking (LK), etc. A common practice of oracles is to examine the opcode and stack content~\cite{cao_survey_2022, kalra_zeus_2018, ma_pluto_2022, luu_making_2016, Groce2021a, Feist2019, wesley_verifying_2022}. For example, \textbf{Unhandled Exceptions (UE)} occur when the contract doesn't check for exceptions after calling external functions. At the opcode level, this is detected by identifying opcodes like \texttt{DELEGATECALL} that aren't surrounded by proper error handling. \textbf{Re-entrance (RE)} vulnerabilities occur when a function can be re-entered before its previous execution completes, potentially causing inconsistent states or fund drains. This is detectable by spotting external calls (\texttt{CALL}, \texttt{DELEGATECALL}, etc.) followed by state-altering opcodes like \texttt{SSTORE}. \textbf{Suicidal (SC)} contracts can self-destruct unexpectedly, leading to data or fund loss. This is identified by the \texttt{SELFDESTRUCT} opcode, especially when it's reachable via external calls. \textbf{Block State Dependency (BD)} happens when a contract's logic relies on predictable block properties like \texttt{BLOCKHASH} or \texttt{TIMESTAMP}, making it vulnerable to manipulation. \textbf{Ether Leaking (LK)} vulnerabilities involve contracts allowing arbitrary users to retrieve ether from the contract, mainly detected by the Ethereum Virtual Machine (EVM) ether-related interface. All these oracles are practical for detecting underlying vulnerabilities. However, they are unable to detect MEVuls due to their logic-based nature. As a consequence, traditional oracles report many unrelated underlying vulnerabilities but miss the real danger lying in monetarily damaging exploitations.


\subsection{Motivation Example}\label{sec:back_motiv}

We use an example of monetarily damaging exploitation to demonstrate the limitations of existing fuzzing tools and the motivation behind our method.

\begin{figure}[ht]
\centering
\includegraphics[width=0.47\textwidth]{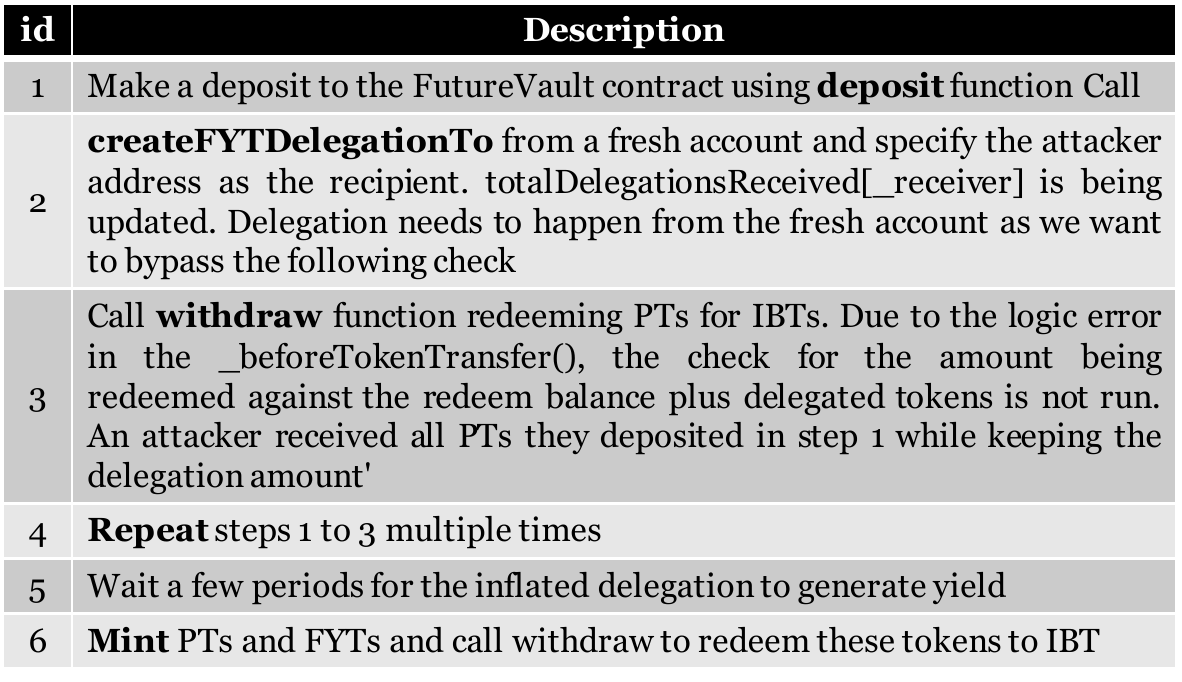}
\caption{The PoC Example in an Audit Report\cite{immunefi_silo_2023}}
  \label{fig:article}
\end{figure}

\PP{Monetarily Damaging Exploitation}
As the number of smart contracts soared, more MEVuls causing huge financial losses have been reported through the bug bounty platform. Proof of Concept (PoC) is a simplified form of vulnerability report, a step-by-step guide showing how the vulnerability is exploited. PoC is practical because developers can rigorously follow it to interact with the smart contract, reproduce the MEVul, and briefly reveal the most crucial exploitations. \autoref{fig:article} is a PoC example reported on an audit website~\cite{immunefi_website}. In the description, \autoref{fig:article} requires calling more than 5 functions with corresponding parameters in the correct order. As a result of execution, the attacker successfully withdrew a large amount of tokens illegally from the benign contract. This PoC showcases a critical vulnerability with potentially severe consequences. However, existing tools are ineffective in addressing this issue from both function generation strategy and oracle perspective.

\PP{Function Generation Strategy's Failure}
Current fuzzing tools use a random strategy to generate the vulnerable function sequence. Some adanved tools support directed fuzzing, meaning that the generated functions are directed by machine learning or deep learning models~\cite{sendner_smarter_2023, he_learning_2019,abdelaziz_smart_2023, Choi2021, godefroid_learnfuzz_2017, jiang_contractfuzzer_2018, liu_rethinking_2023, zhang_deltafuzz_2022,kolosnjaji_empowering_2017, she_neuzz_2019}. However, they struggle with generating the complicated function sequences to trigger MEVul like in PoC of \autoref{fig:article}. The reason is that the learning resources of these models are mainly static tools, which only focus on traditional opcode-level vulnerabilities, so they are efficient in searching traditional vulnerabilities but terrible at searching MEVuls.

\noindent \textit{Solution by \sys.}
We propose that the \textsc{FAuditor} should be directed by auditors' experiences but not by the results of existing tools. The auditors' experiences, such as PoC, auditing articles, and reports, describe the code's higher-level logic, and they target exploitations, which are more informative for searching MEVuls. For example, in the PoC of \autoref{fig:article}, which requires function sequence: 
\begin{align*}
  &\texttt{deposit()} \rightarrow \texttt{createFYTDelegationTo()} \rightarrow \\
  &\texttt{withdraw()} \rightarrow \texttt{LOOP 1-3} \rightarrow \texttt{mint()} \rightarrow \texttt{withdraw()}
\end{align*}
If \textsc{FAuditor} has learned this PoC and starts to analyze another similar contract, it can generate a similar sequence:
\begin{align*}
  &\texttt{save()} \rightarrow \texttt{delegateTo()} \rightarrow \texttt{withdraw()} \rightarrow \\
  &\texttt{LOOP 1-3} \rightarrow \texttt{mint()} \rightarrow \texttt{withdraw()}
\end{align*}
In this sequence, the MEVul is more likely to be triggered when the targeted contract‘s business logic is similar to the previously learned one.


\PP{Traditional Oracle's Failure}
The traditional oracles are hardly aware of the features of MEVul. In \autoref{fig:article}, the attacker unrestrictedly \texttt{withdraw} the tokens from the contract, and the root reason is a direct \texttt{deposit} made at the first transaction, causing the swap rate of the token to explode. It is the deficiency of the code logic causing financial loss, while traditional oracles only check opcode patterns or some EVM global variables and are not aware of the important variables such as token price and account balance in this exploitation. Thus, they are not capable of detecting MEVuls.

\noindent \textit{Solution by \sys.}
\textsc{FAuditor} employs key variables like \texttt{account}, \texttt{balance}, \texttt{token price}, \texttt{exchange rate}, etc., to capture the critical feature of MEVul. In the example of \autoref{fig:article}, because \textsc{FAuditor} is monitoring the swap rate of the token (\texttt{PT}) in real-time, the drastic fluctuation of the rate will be captured as a vital MEVul.

\section{The Concept of \textsc{FAuditor}}

\begin{figure*}[ht]
  \centering
  \includegraphics[width=\textwidth]{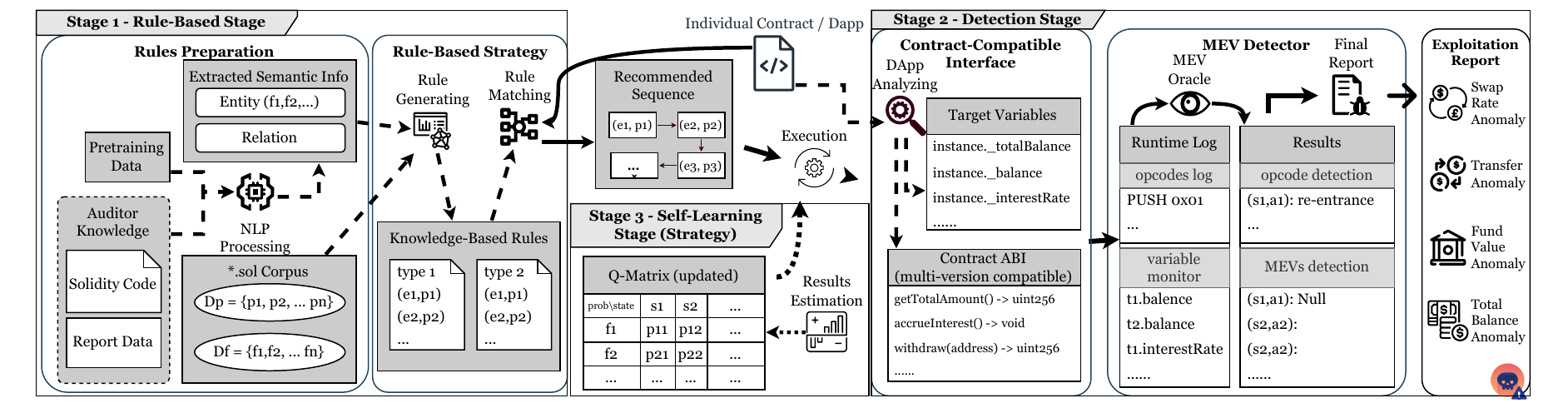}
  \caption{\textbf{The Overview of \textsc{FAuditor}.} \textit{We use different arrow types to present different system workflows. (i) - -: dotted segment arrow represents the data/interface preparation process, (ii) --: full line arrow represents the main detection process, and (iii) ··: dotted arrow represents the feedback process for self-learning.}}\label{fig:system_framework}
\end{figure*}


In this paper, we address the gap in detecting monetarily exploitable vulnerabilities (MEVuls) with \textsc{FAuditor}, a fuzzing framework. This section outlines \textsc{FAuditor}'s workflow, challenges, and techniques.

\subsection{Workflow of FAuditor}\label{sec:overview_framework}

\autoref{fig:system_framework} shows the workflow of \sys, divided into three stages: rule-based, detection, and self-learning. The system inputs smart contracts and outputs exploitation reports to identify MEVuls.

\PP{Rule-Based Stage} This stage uses auditor-derived rules to generate function sequences targeting MEVuls. A pre-trained NLP model extracts key semantics (e.g., function names and sequences) from auditor reports, which are then used to match specific rules to contract types, guiding the analysis.

\PP{Detection Stage} In this stage, contract execution is analyzed for MEVuls. The Contract-Compatible Interface resolves version dependencies and identifies proxy-deployed contracts. The MEVul Detector checks key variables based on formalized MEVul concepts, classifying vulnerabilities into four categories: Transfer Arrival, Balance Conservation, Swap Rate Stability, and Fund Value Stability.

\PP{Self-Learning Stage} This stage refines the function generation process by learning from previous detection outcomes. The Self-Learning Strategy dynamically adjusts function sequences based on feedback, gradually optimizing and replacing the rule-based method for future test case generation.

\subsection{Challenges and Techniques}
Before showing the design of \sys, we discuss the technical challenges (C) of realizing \sys and the corresponding techniques (T) to address these challenges.

\PP{C1 - Precise Definition and Detection for MEVuls}
Existing fuzzing methods have primarily focused on identifying underlying vulnerabilities and overlooked economic exploitability, leading to a lack of a clear definition of MEVul. In terms of detection, the traditional oracles of underlying vulnerabilities are incapable of detecting MEVuls, while some invariant expression techniques are specific to particular contracts and not easily applicable across the board.

\PP{T1 - MEVul Formalization and Automatic Detector}
First, to identify the MEVuls, we formalize their definitions based on smart contracts' interfaces in section \autoref{sec_vul_def} into four vulnerability types strongly correlated to financial loss. These generic categories can be applied to decentralized applications (DApp) involving interactions between multi-contracts/tokens. Second, to create a feasible MEVul detector across different contracts, we automated the detector to recognize and extract the key variable's value from smart contracts, which is implemented through the ABI selector instrumented in the Contract Compatibility Interface. Consequently, the \textit{MEVul Detector} can efficiently identify the MEVul.

\PP{C2 - Practical Search Strategy}
Current search strategies employing function generation are designed for underlying vulnerabilities such as re-entrance and suicidal behaviors. Thus, these strategies are too time-consuming and inaccurate for the purpose of MEVuls. Our assessment of existing tools in \autoref{sec:exp_RQ1_replicate} and \autoref{sec:exp_RQ2_novel} showed that they either have a high rate of false positives or a significant number of timeouts when attempting to replicate existing MEVuls.

\PP{T2 - Rule-Based Strategy and Self-Learning Strategy}
First, to design a basic strategy for MEVul, we collected auditing reports from skilled auditors to create a knowledge-based dataset. The dataset comprises the PoC of exploitation and corresponding codebases. Second, we extract rules from this dataset and then enable the model to generate high-risk function sequences based on these extracted rules and the program's running status. This \textit{Rule-Based Strategy} is defined in \autoref{sec_brain_def}.
Third, to enhance the usability of existing rules, we developed the Self-Learning Strategy. This strategy utilizes the fuzzing history as feedback to optimize the function generation procedure. The \textit{Self-Learning Strategy} is defined in \autoref{sec_fuzzing_self_study}.


\section{The Design of FAuditor}\label{sec:fuzzer_strategy}

This section presents the design of \textsc{FAuditor}. It covers the definition of the Rule-Based Strategy, which generates function sequences using rules extracted from auditor reports; the definition of MEVul oracle, which provides expressions for four types of monetarily exploitable vulnerabilities (MEVuls); and the definition of the Self-Learning Strategy, which learns from fuzzing history based on a Markov process to optimize subsequent function generation procedure.

\subsection{Rule-Based Strategy}\label{sec_brain_def}
In this section, we define the Rule-Based Strategy, which uses rules extracted from auditor reports to generate sequences of functions.

\PP{Preliminary}
For preliminary purposes, let's define the Fuzzing Strategy as a random variable that represents the current state of the fuzzer and the next transaction to be created:
\begin{definition}[Fuzzing Strategy]\label{def_fuzzing_strategy}
  A Fuzzing Strategy is a random variable presented by $\pi: \mathcal{T} \times \mathcal{T}^{*} \rightarrow \mathcal{T}$, where $\mathcal{T}$ is the collection of transaction $t$, $\mathcal{T}^*$ is the set of transactions sequence $\vec{t}:= (t_1, t_2, \dots, t_n)$.
\end{definition}
$\pi$ is a random variable mapping from the Cartesian product of the transaction and its sequence space to another transaction space. So, knowing the existing transaction sequence generated from contract $C$, the random variable of the next transaction is $\pi: \mathcal{T}_C \times \mathcal{T}^{*}_{C} \rightarrow \mathcal{T}_C$. $\pi$ differs in different strategies.

\PP{Procedure of the Rule-Based Strategy}
This module executes a coercive Rule-Based Strategy as the first stage of the \textsc{FAuditor}'s Fuzzing Strategy when the fuzzing cold starts. The strategy is to make the fuzzer generate the function sequence strictly following the rules constructed from the auditor reports.

Firstly, we define the rule, which is based on the definition of transaction and its power set:
\begin{definition}[Rule of Strategy]\label{def_rule}
  When $\mathcal{T}_r$ is a set of transactions, and $\mathcal{T}^*_r$ is the power set of it. A rule of strategy is denoted as a transaction sequence $\vec{t_r}$, that
\begin{equation}
  \vec{t_{r}} =(t^{'}_1,t^{'}_2,\dots,t^{'}_n),\qquad
  \begin{aligned}
    & \vec{t_r} \in \mathcal{T}^*_r \\
    & t^{'}_i \in \mathcal{T}_r, 0<i \leq n 
  \end{aligned}
\end{equation}
\end{definition}
Having the rule $\vec{t_r}$, and assuming $\vec{t}=(t_1,t_2,\dots,t_i)$ is the sequence previously generated from a certain contract, the next transaction $t_{i+1}$ will be generated according to the following procedure:

When $i=0$, the strategy searches for the rule fit to this contract:
\begin{equation}
  \pi(t_{i+1}\mid (t_i,\vec{t_r})) \sim \mathcal{U}_{t_{i+1} \in \mathcal{T}_C}
  \label{eq_aveSim}
\end{equation}
It keeps randomizing with a random uniform distribution, with $i$ remaining at 0. Once $t_0 \in \mathcal{T}_r$, the fuzzing process is trapped into the Rule-Based Strategy, and $i$ is set to 1.

When $ i \neq 0$, the strategy is to make the fuzzing process follow the rule strictly. If $\exists t_{i+1} \in \mathcal{T}_C, t_{i+1} = t^{'}_{i+1} $:
\begin{equation}
  \pi(t_{i+1}\mid (t_i,\vec{t_r})) \sim \mathbb{P}(t_{i+1}=t^{'}_{i+1}) = 1
\end{equation}
, while if $\not \exists t_{i+1} \in \mathcal{T}_C, t_{i+1} = t^{'}_{i+1} $:
\begin{equation}
  \pi(t_{i+1}\mid (t_i,\vec{t_r})) \sim \mathcal{U}_{t_{i+1} \in \mathcal{T}_C}
\end{equation}
And at the end of the rule, where $i = n$, we reset $i:=1$ so the fuzzing process restarts.

\subsection{MEVul Oracle}\label{sec_vul_def}

In this section, we define the MEVul-related concepts and MEVul oracles. MEVuls are vulnerabilities directly related to monetarily damaging exploitation. Based on our summary of prevalent exploitations, we categorized them into four security paradigms: Transfer Arrival, Total Balance Conservation, Swap Rate Stability, and Fund Value Stability. 

\PP{Preliminary}
First, we define some useful variables, which will help describe our paradigms later. Having the contract $C$ deployed, its function collection is represented as $\mathcal{F}$, and $\mathcal{F}_p \subseteq \mathcal{F}$ is its payable functions collection. For each $f \in \mathcal{F}_p$, it takes (i) $a \in \mathcal{A}$ -- address from address collection, (ii) $v \in \mathbb{R}^* $ -- transfer amount, and (iii) $k \in \mathcal{K}$ -- the type of token, as parameters. Besides, every contract maintains a table of users' balance $B: \mathcal{A} \times \mathcal{K} \rightarrow \mathbb{R}^*$. For example, $B(a, k)$ = $v$ means that for token $k$, the balance of account $a$ is $v$. 

We then define the token's price and the swap rate between tokens to better describe their relationship.
\begin{definition}[Token Price]
Let $p: \mathcal{K} \rightarrow \mathbb{R}^*$ be a function representing the token price $k$. For a given token $k \in \mathcal{K}$, it has $p(k) = v, v \in \mathbb{R}^*$, which means the price of one token $k$ is $v$.
\end{definition}
\begin{definition}[Token Swap Rate]
Let $e: \mathcal{K} \times \mathcal{K} \rightarrow \mathbb{R}^*$ be a function representing the swap rate between tokens $k_i$ and $k_j$. For any pair of tokens $(k_i, k_j) \in \mathcal{K} \times \mathcal{K}$, it has $v_{ij} = e(k_i, k_j), v_{ij} \in \mathbb{R}^*$, and $ p(k_i) v_{ij}= p(k_j)$.
\end{definition}
It means the swap rate between token $k_i$ and $k_j$ is $v_{ij} = e(k_i, k_j)$, and we can pay $v_{ij}$ number of token $k_i$ to get one token $k_j$. Both prices and swap rates are being dynamically changed by transactions made on the blockchain. We express it as: 
\begin{equation}
  (p_i,e_i)\xrightarrow{\vec{t}} (p_{i+1},e_{i+1})
\end{equation}
, where $\vec{t}$ is a sequence of transactions made on the blockchain, and $p_i, e_i$ are the states of price and swap rate at time $i$.

The same dynamic property is true for the states of balance. When we use $B_i$ to illustrate the balance states of the contract, each payable function call will cause the transfer of the state: $B_i \xrightarrow{\vec{t}} B_{i+1}$. In the case of $\vec{t}$ containing only one single transaction, we also use the function $f(\vec{a}_\text{from}, \vec{a}_\text{to},\vec{k},\vec{v})$ to express the state transfer: 
\begin{equation}
  B_i \xrightarrow{f(\vec{a}_\text{from}, \vec{a}_\text{to},\vec{k},\vec{v})}  B_{i+1}
\end{equation}

\PP{Paradigm - Transfer Arrival}
Transfer Arrival means the money is safely transferred from the starter account to the target account and is not stolen by the malicious account. It is the most basic security paradigm covering cases like token theft, token self-destructs, etc.

We divide users' accounts into two different types: the benign account and the malicious account, expressed as $\mathcal{A}_n, \mathcal{A}_m \subset \mathcal{A}$. Transfer Arrival is defined as:
\begin{proposition}[Transfer Arrival]\label{propo_transfer_arrival}
  A transaction is a Transfer Arrival when $B_i \xrightarrow{f(a_{\text{from}}, a_{\text{to}}, k, v)} B_{i+1}$ with $f \in \mathcal{T}_p, a_{\text{from}}, a_{\text{to}} \in \mathcal{A}_n$; it satisfies:
  \begin{equation}
    \begin{aligned}
      &B_{i}(a_{\text{from}}, k)-B_{i+1}(a_{\text{from}}, k)
      = B_{i+1}(a_{\text{to}}, k)-B_i(a_{\text{to}}, k) 
      = 2v
    \end{aligned}
  \end{equation}
  , and for $\forall a_m \in \mathcal{A}_m$, it has:
  \begin{equation}
    B_{i}(a_m, k) = B_{i+1}(a_m, k).
  \end{equation}
\end{proposition}
This paradigm ensures that the difference in the balance of the two benign accounts equals the transfer amount, and the balance of the malicious account is unchanged. When the gas fee is not accounted for, it means the money is transferred safely from the starter account to the target account, not being stolen by the malicious account.

\PP{Paradigm - Total Balance Conservation}
Total Balance Conservation is another basic security paradigm, which means the total value held by the benign accounts is conserved during the transaction. Total Balance Conservation is defined as:
\begin{proposition}[Total Balance Conservation]\label{propo_balance_cons}
  Having $f$ as the function call made in the transaction, $p_i(k)$ as the price of token $k$ at time $i$, the transaction maintains Total Balance Conservation, when $B_i \xrightarrow{f(\cdot)} B_{i+1}$, $f \in \mathcal{F}_p$:
  \begin{equation}
    \sum_{a \in \mathcal{A}_n, k \in \mathcal{K}}B_i(a, k)p_i(k) = \sum_{a \in \mathcal{A}_n, k \in \mathcal{K}}B_{i+1}(a, k)p_i(k)
  \end{equation}
  , where $\mathcal{A}_n$ is the set of benign accounts, $\mathcal{K}$ is the set of tokens.
\end{proposition}
The accounts above include both the user account and the contract account. This definition ensures that the total value held by the whole group of benign accounts will not disappear or decrease. Its difference from Transfer Arrival is that Total Balance Conservation aims to check the anomaly of the total value of the entire group rather than individual accounts.

\PP{Paradigm - Swap Rate Stability}\label{sec_vul_def_interest_rate}
Swap Rate Stability is a high-level security measure that ensures the token swap rate remains unmanipulated. The swap rate is a crucial factor in the field of decentralized finance (DeFi), directly impacting the token's value. Manipulating the swap rate poses a significant vulnerability in the DeFi space, and if it occurs, it can result in substantial financial losses and potentially lead to the project's failure.

We utilize the on-chain price of the token to specify the definition. The on-chain price of a token is expressed as $p_{\text{obs}}(k) \in \mathbb{R^*}$, meaning the observed on-chain price of token $k$. And a tolerance $\epsilon \in [0,1]$ parameter being tuned artificially for practicability. The Rate Stability is defined as:
\begin{proposition}[Swap Rate Stability]\label{propos_rate_stability}
  Having $p_i \xrightarrow{\vec{t}} p_{i+1}$, $\vec{t} \in \mathcal{T}_p$, $\vec{t}$ is the sequence of transactions made on the blockchain, $p_i(k)$ is the under-test price of token $k$ at time $i$, $p_{\text{obs}}(k)$ is the observed on-chain price of token $k$ at time $i$, and $\epsilon \in [0,1]$ is the tolerance parameter. For $\forall k \in \mathcal{K}, \forall i \in \mathbb{N}$, we define the Swap Rate Stability as:
  \begin{equation}
    \begin{aligned}
      \left| \frac{p_{i}(k)}{p_{\text{obs}}(k)} - 1 \right| \leq \epsilon &\to \left| \frac{p_{i+1}(k)}{p_{\text{obs}}(k)} - 1 \right| \leq \epsilon \\
      & \text{and} \\
      \left| \frac{p_{i}(k)}{p_{\text{i+1}}(k)} - 1 \right| \leq \epsilon &\to \left| \frac{p_{i+1}(k)}{p_{\text{i+2}}(k)} - 1 \right| \leq \epsilon. \\
    \end{aligned}
  \end{equation}
\end{proposition}
This definition describes a scenario in which the current token price is near the observed price, and the subsequent token price should also be close to the observed price. Additionally, price changes should be stable and not exceed the tolerance value of $\epsilon$. This definition effectively addresses instances of abnormal price fluctuations, typically caused by malicious manipulation.

\PP{Paradigm - Fund Value Stability}
Fund Value Stability is another high-level security paradigm, which means the contract's total price is stable. The total price is the sum of all the contract's token prices. A significant fluctuation in the total price usually means the contract is unsafe and probably under attack.

To specify the definition of Fund Value Stability, we introduce the coefficient $\lambda \in [0,1]$, which is a parameter that describes the fluctuation level of the total value.
\begin{proposition}[Fund Value Stability]\label{propos_funding_stability}
  Having $B_0 \xrightarrow{\vec{t}} B_n$, $\vec{t} \in \mathcal{T}^*_p$, we have total value of the contract at certain time $i, i \in [1,\infty)$, while the initial value is defined by the observed on-chain price $p_{obs}(k)$, they are expressed as $v_0, v_i$, $v_0, v_i \in \mathbb{R}^*$:
  \begin{equation}
    \begin{aligned}
      v_i = \sum_{a \in \mathcal{A}, k \in \mathcal{K}}B_i(a, k)p_{i}(k),
      v_0 = \sum_{a \in \mathcal{A}, k \in \mathcal{K}}B_0(a, k)p_{obs}(k)
    \end{aligned}
  \end{equation}
  , given the fluctuation coefficient rate $\lambda \in [0,1]$, we define the Funding Stability as:
  \begin{equation}
    v_0(1+\lambda) \geq v_i \geq v_0(1-\lambda)
  \end{equation}
\end{proposition}
This definition describes the contract's total value as being consistent with the initial value. Typically, for regular financial operations like borrowing and lending, collateral should be required to guarantee the financial security of the fund. This ensures that the contract's total value remains stable. If the stability is compromised, it means that the contract's value has been manipulated or stolen.


\subsection{Self-Learning Strategy }\label{sec_fuzzing_self_study}

In this section, we define the Self-Learning Strategy and related concepts. The self-learning strategy is based on a Markov process, using fuzzing history to optimize the function generation procedure.

\PP{Preliminary I - Markov Strategy}\label{sec_markov}
The Markov strategy is the basis for the Self-Learning Strategy. It involves the fuzzer generating a sequence according to the Markov process.

The current state of the process is simplified as $t \in \mathcal{T}_C$, temporarily assuming the Markov state only depends on the current transaction. The next transaction is represented by a random variable that maps from the transaction space to another $\pi: \mathcal{T}_C  \rightarrow \mathcal{T}_C$.

A random Markov strategy randomly selects the next transaction without considering the current transaction.
\begin{equation}
  \pi(t_{i+1} | t_{i}) \sim \mathcal{U}_{t_{i+1} \in \mathcal{T}_C}
  \label{eq_markovSim}
\end{equation}
The difference between the Markov strategy and the Rules-Based Strategy is that \textbf{the selection of the next function is not decided by the knowledge-based rule anymore, but by the current state of the fuzzing process.} The conditional probability selection expression is $\pi(t_{i+1} | t_{i})$ rather than $\pi(t_{i+1}\mid (t_i,\vec{t_r}))$.

To construct a practicable Markov strategy, we replace its probability distribution to improve its capability to discover bugs. For simplification, We use the $\text{Policy}(t_{i})$ to express the new probability distribution:
\begin{equation}\label{eq:policy_incomplete}
  \pi(t_{i+1} | t_{i}) \sim \text{Policy}(t_{i})
\end{equation}
And we will give a detailed definition of the $\text{Policy}(t_{i})$ sooner.

\PP{Preliminary II - Self-Learning Policy}
For preliminary purposes, we define the concepts related to the self-learning policy, including Fuzzing State, Reward Function, and Quality Function.

We define the execution state of the fuzzer, named the Fuzzing State, which records the useful information of the fuzzing process, helping the fuzzer to make the subsequent decision.
\begin{definition}[Fuzzing State]
  The Fuzzing State is:
  \begin{equation}
    \begin{aligned}
      \mathcal{S} = \{&s \mid s = (t, e, L, m), \forall t \in \mathcal{T}, e \in \mathcal{E}, L \in \mathcal{P}(\mathcal{E}), m \in \mathcal{M}\}
    \end{aligned}
  \end{equation}
  , where $t \in \mathcal{T}$ is the current transaction, $e \in \mathcal{E}$ is error type currently detected, $L \in \mathcal{P}(\mathcal{E})$ is the labeled errors of current contract, $\mathcal{P}(\mathcal{E})$ is the power set of $\mathcal{E}$. And $m \in \{0,1/2,1\}$ is the Measurement of the state. The Measurement $m$ is specified as:
  \begin{equation}
    m = 
    \left\{
      \begin{aligned}
        &1, && L \neq \emptyset \wedge e\in L \\
        &1/2, && L = \emptyset \wedge e \neq \emptyset \\
        &0, && \text{otherwise}
      \end{aligned}
    \right.
  \end{equation}
\end{definition}
Specifically, the label $L$ includes MEVuls
and the other traditional types
of vulnerabilities. Generally speaking, the Fuzzing State mainly presents the performance of the fuzzer on the current contract, where the $m$ returns the positive feedback when the correct vulnerability type is identified or a new vulnerability is detected.

Having the Fuzzing State definition, we define the reward function. The reward function is to quantify the measurement of the state and feedback reward value to the fuzzing process:
\begin{definition}[Reward Function]
  $s_i, s_{i+1} \in \mathcal{S}$ are two consecutive states in a fuzzing process, $s_n \in \mathcal{S}$ is the last state in one fuzzing process. If $i<n-1$, The Reward Function is:
  \begin{equation}
    r_{i+1} = 
    \left\{
      \begin{aligned}
        & 1, && m_{i+1} \leq m_i \\
        & 0, && \text{otherwise}\\
      \end{aligned}
      \right.
  \end{equation}
\end{definition}
The reward function is calculated based on the measurement of the current state compared to the previous state; if the current state is better, the reward is 1; otherwise, it is 0.

We then define the Quality Function, which evaluates the quality of every potential selection for the next step. It should reflect all feedback from previous states.
\begin{definition}[$Q$-Function]\label{def_qFunc}
  Having Fuzzing State $s$ and the function $f$ being called in the current step, $s \in \mathcal{S}, f \in \mathcal{F}$, the Quality Function is:
  \begin{equation}
    Q_{n+1}(s, f) =
    \left\{
      \begin{aligned}
        &1, && i=0 \\
        &\frac{\sum_{i\in (0,n)}r_{i}(s,f)\cdot  m_{i}(s,f)}{n}, && i\neq 0
      \end{aligned}
    \right.
  \end{equation}
\end{definition}
$Q$-Function is the average of the previous summation of $r\cdot m$, which reflects the quality of the action on the current contract for the current state. Therefore, for the next step, the fuzzer should select the function $f\in \mathcal{F}$ that maximizes the $Q$-Function because it is the strategy that is most likely to trigger the vulnerabilities.

For recursive programming, we should deduce the recurrence expression of the $Q$-Function from the \autoref{def_qFunc}:
\begin{theorem}[Recursive $Q$-Function]
  Having current Rewared Function $r_n$, Measurement $m_n$, and the $Q$-Function $Q_{n}$, the recurrence expression is:
  \begin{equation}
    Q_{n+1} = \frac{(n-1)}{n} Q_{n} + \frac{r_n m_n}{n}
  \end{equation}
\end{theorem}

Having all the definitions above, we can complete the definition of $\text{Policy}(\cdot)$ not finished in \autoref{eq:policy_incomplete}. Firstly, we define the Greedy Policy, which is the Policy that always chooses the action that maximizes the $Q$-Function:

\begin{proposition}[Greedy Policy]
  Having the current state $s_i \in \mathcal{S}$, the current function $f_i \in \mathcal{F}$, the next function $f_{i+1} \in \mathcal{F}$ to be taken, $\text{Policy}: S \times \mathcal{F} \rightarrow [0,1]$ is defined as a probability distribution mapping from the state space and the function space to the probability space:
  \begin{equation}
    \text{Policy}(s_i) : \mathbb{P}(f_{i+1}=\operatorname*{arg\,max}_{f \in \mathcal{F}} Q(s_i,f_i)) = 1
  \end{equation}

\end{proposition}
The Greedy Policy is the most straightforward, always choosing the action that maximizes the $Q$-Function. 

However, the Greedy Policy is impractical in real-world situations because it may cause the fuzzer to be trapped in the local optimum. To avoid this, we introduce the $\epsilon$-Policy, which is a policy that always chooses the action that maximizes the $Q$-Function with a probability of $1-\epsilon$ and chooses the other actions with a probability of $\epsilon/n$:
\begin{proposition}[$\epsilon$-Policy]
  Having the function $\bar{f}$ maximizing the $Q$-Function:
  \begin{equation}
    \bar{f} = \operatorname*{arg\,max}_{f \in \mathcal{F}} Q(s_i,f_i)
  \end{equation}
  , the definition of the $\epsilon$-policy is:
  \begin{equation}
    \text{Policy}_{\epsilon}(s_i) : \mathbb{P}(f_{i+1}\mid s_i) = 
    \left\{
    \begin{aligned}
      &1-\epsilon, && f_{i+1} = \bar{f} \\
      &\epsilon/n, && f_{i+1} \in \mathcal{F} \setminus \bar{f} \\
      & 0, && \text{otherwise}
    \end{aligned}
    \right.
  \end{equation}
\end{proposition}

\PP{Procedure of the Self-Learning Strategy}
The Self-Learning Strategy, based on the previous Markov strategy, is the secondary function generation strategy to be executed after the Rule-Based Strategy. The Strategy's goal is to learn from the fuzzing history and make decisions based on the current state of the process.

The Self-Learning Strategy is a Markov process-based strategy with $\epsilon$-Policy:
\begin{proposition}[Self-Learning Strategy]
  The state of the Self-Learning Strategy is:
  \begin{equation}
    \begin{aligned}
      \mathcal{S} = &\{s \mid s = (t, e, L, m), \forall t \in \mathcal{T}, e \in \mathcal{E}, L \in \mathcal{P}(\mathcal{E}), m \in \mathcal{M}\}
    \end{aligned}
  \end{equation}
  And having $\text{Policy}_{\epsilon}$, the State Transition Variable of the Self-Learning Strategy is defined as:
  \begin{equation}
    \pi(s_{i+1} \mid s_i) \sim \text{Policy}_{\epsilon}(s_i)
  \end{equation}
\end{proposition}
In conclusion, the Self-Learning Strategy utilizes the current Fuzzing State and detection results from the previous round of fuzzing as inputs to generate the next high-risk function.

\section{Implementation of FAuditor}\label{sec:implementation}

This section briefly outlines the implementation of \textsc{FAuditor}, as shown in \autoref{fig:system_framework}. 

\PP{Rule Preparation}
Rule preparation involves transforming collected text data into structured rules for the Rule-Based Strategy through three steps: NLP model pre-training, semantic extraction, and corpus construction. (i) NLP Model Pre-training: We fine-tune an NLP model on the SciERC dataset~\cite{luan_multi-task_2018} to capture scientific relationships, then train it further on a Solidity dataset~\cite{noauthor_smartbugssmartbugs-curated_2024} to understand smart contract-specific terminology. (ii) Semantic Extraction: Auditor reports are analyzed to extract entities and their relationships, providing valuable context for rule generation. (iii) Corpus Construction: A manually curated corpus of Solidity function names, parameters, and relationships ensures accurate rule generation for vulnerability detection.


\PP{Strategy Training}
As a training resource, we collected codebases and corresponding auditing reports from 441 contracts across various auditing platforms, including Immunefi\cite{immunefi_website}, OpenZeppelin~\cite{wesley_verifying_2022}, PeckShield~\cite{noauthor_uniswaplendfme_nodate}, and Code4Rena~\cite{noauthor_code4rena_nodate}. These reports, containing PoCs and codebases, were used to refine both the Rule-Based and Self-Learning strategies. The dataset is not restricted to Monetarily Exploitable Vulnerabilities (MEVuls) but encompasses a broad range of vulnerabilities. To optimize model performance, we applied 5-fold cross-validation to ensure robust model selection. Notably, this training dataset is excluded from the evaluation phase (\autoref{sec:experiment}) to prevent data leakage and mitigate overfitting, ensuring that the reported performance reflects the model's ability to generalize to unseen data.

After training, the Rule-Based Strategy generates transaction sequences based on predefined rules. It extracts key rules from auditor reports and matches them to the target contract's functions to recommend the next transaction for analysis. The Self-Learning Strategy improves fuzzing by learning from previous runs, focusing on functions more likely to expose vulnerabilities. It dynamically adjusts the strategy to balance exploration and exploitation based on historical results.



\PP{MEVul Detector}\label{sec:eye_implementation}
The MEVul Detector identifies vulnerabilities in smart contracts, supporting both traditional and MEVul-specific oracles. Detection involves (i) Extracting ABI information via the Contract-Compatible Interface. (ii) Using MEVul Oracles to detect anomalies based on key variables extracted from ABI functions, such as $\texttt{balanceOf()}\rightarrow B$ and $\texttt{getPrice()}\rightarrow p$. Each detector uses specific variables to assess vulnerabilities, and their mapping relations can be expressed as (I) $\frac{\texttt{balanceOf()}}{TA}$, (II) $\frac{\texttt{getPrice()},\texttt{balanceOf()}}{BC}$, (III) $\frac{\texttt{getPrice()},\texttt{getAmountOut()}}{ES}$, and (IV) $\frac{\texttt{balanceOf()}, \texttt{getPrice()},\texttt{getAmountOut()}}{VS}$. 

\PP{Contract-Compatible Interface} This interface addresses Dapp-related challenges, including EVM Version Compatibility and Proxy-deployed Contract Discovery. (i) EVM Version Compatibility incorporates a contract hash discovery function to support multiple EVM versions and swarm hashes in Dapp testing. (ii) Proxy Pattern Contract Rediscovery integrates a Contract Mocker to detect and interact with proxy-deployed contracts, ensuring seamless interaction.

\section{Evaluation}\label{sec:experiment}

To evaluate \sys, we conducted large-scale experiments with over 5,760 CPU hours. This evaluation aims to answer the following four research questions:

\vspace{-\parskip} \noindent \textbf{RQ1:} How many existing monetarily exploitable vulnerabilities (MEVuls) can \sys reproduce compared to state-of-the-art fuzzers? 

\vspace{-\parskip} \noindent \textbf{RQ2:} How many zero-day MEVuls can \textsc{FAuditor} discover?

\vspace{-\parskip} \noindent \textbf{RQ3:} To what extent does \sys achieve higher code coverage compared to state-of-the-art fuzzers?

\vspace{-\parskip} \noindent \textbf{RQ4:} To what extent does \sys's new strategy reduce the time required to detect vulnerabilities?



\PP{Experimental Setup}\label{sec:exp_setup}
The experiment was conducted on a desktop machine with the following configuration: Intel 12th Gen i5-12600K CPU (3.70GHz) 32GB RAM for fuzzing task, Nvidia RTX 3070Ti GPU for strategy learning task, all running on Ubuntu 20.04 LTS OS with WSL 2.0 on Windows 11. The method is implemented in Python 3.6, and the main libraries used are brownie, truffle~\cite{noauthor_truffle_nodate}, ganache~\cite{noauthor_ganache_nodate}, and go-ethereum~\cite{noauthor_goeth_nodate}. 


\PP{Benchmark Fuzzer}
We chose Smartian~\cite{Choi2021}, Echidna~\cite{Groce2021a}, ILF~\cite{he_smart_2020}, and ItyFuzzer~\cite{ityfuzz} as benchmark fuzzers for comparison. Echidna~\cite{Groce2021a} is the most widely used, with strong invariant customizability but a naive searching strategy targeting traditional vulnerabilities and violations of user-specified invariants. Smartian~\cite{Choi2021} and ItyFuzzer~\cite{ityfuzz} represent the state-of-the-art symbolic execution-guided fuzzing. ILF~\cite{he_smart_2020} represents the state-of-the-art machine learning-guided fuzzing to iteratively improve test generation by learning from previously discovered vulnerabilities. These tools cover different search strategies for detecting vulnerabilities in smart contracts.


\PP{Datasets}\label{sec:experiment_dataset_benchmark}
The evaluation dataset consists of contracts labeled with vulnerabilities for functional validation and performance evaluation. 
We compiled the benchmark dataset by combining all the datasets used by Echidna~\cite{Groce2021a}, ContractFuzzer~\cite{jiang_contractfuzzer_2018}, and SMS~\cite{SMS2023} for the traditional vulnerability types. 
Next, we collected MEVuls data from auditing platforms such as Immunefi\cite{immunefi_website}, Open-Zeppelin~\cite{wesley_verifying_2022}, PeckShield~\cite{noauthor_uniswaplendfme_nodate} and Code4Rena~\cite{noauthor_code4rena_nodate}, covering the period from 2019 to 2024. 
This dataset was manually curated, adhering to the strict criterion that all reported cases must have led to actual monetary exploitation in real-world scenarios.
Then, these contracts are labeled with MEVul types roughly according to their Proof of Concept (PoC) descriptions.
In total, the overall performance dataset contains 556 labeled contracts, categorized into 418 traditional vulnerability types and 138 MEVul types. The traditional vulnerabilities include 87 Unhandled Exceptions (UE), 80 Re-entrancies (RE), 89 Suicidals (SC), 77 Block State Dependencies (BD), and 86 Leakings (LK). The MEVul types include 42 Transfer Arrival (TA), 32 Total Balance Conservation (BC), 21 Swap/Exchange Rate Stability (ES), and 43 Fund Value Stability (VS).


\subsection{Reproducing Existing MEVuls}\label{sec:exp_RQ1_replicate}
\begin{table*}[ht]
\scriptsize
\rowcolors{2}{gray!10}{gray!30}
\resizebox{\linewidth}{!}{
  \centering
  \begin{tabular}{c c c c c c c c c c c}
  \hline
  \textbf{tx} & \textbf{Instance} & \textbf{Method} & \textbf{Parameters} & \textbf{from} & \textbf{Inner price for} $(t_A,t_B,t_T)$ & $B(t_A)$ & $B(t_B)$ & $B(t_T)$ & $v_{\texttt{internal}}$ & $v_{\texttt{obs}}$\\
  \hline
  1 & TokenA & approve & addr(Protocol), 0.1 & attackers[0]  & (0, 7e-2, 1.1) & 0 & 1428 & 90.9 & 200 & 200 \\
  2 & Protocol & deposit & addr(TokenA), 0.1, false & attackers[0] & (9.99e-1, 7e-2, 1.1) & 0.1 & 1428 & 90.9 & 200.1 & 200.1\\
  3 & TokenA & \textbf{transfer} & addr(Protocol), 1 & attackers[0] & (9.99e-1, 7e-2, 1.1) & 100.1 & 1428 & 90.9 & 300.1 & 200.1\\
  4 & TokenB & approve & addr(Protocol), 100 & attackers[1] & (9.99e-1, 7e-2, 1.1) & 100.1 & 1428 & 90.9 & 300.1 & 200.1\\
  5 & Protocol & deposit & addr(TokenB), 100, true & attackers[1] & (9.99e-1, 7e-2, 1.1) & 100.1 & 2846 & 90.9 & 300.1 & 200.1\\
  6 & Protocol & borrow & addr(TokenA), 1 & attackers[1] & \textbf{(999, 7e-2, 1.1)} & 0.1 & 2846 & 90.9 & \textbf{300.1} & 200.1\\
  7 & Protocol & borrow & addr(Target), 100 & attackers[0] & (999, 7e-2, 1.1) & 0.1 & 2846 & 0 & \textbf{200.1} & 200.1\\
  8 & Protocol & withdraw & addr(TokenB), 100 & attackers[1] & (999, 7e-2, 1.1) & 0.1 & 1428 & 0 & \textbf{100.1} & 200.1\\
  9 & Protocol & liquidation & - & owner & (9.99e-1, 7e-2, 1.1) & 0.2 & 1428 & 0 & \textbf{100.2} & 200.1\\
  \hline
  \end{tabular}
}
  \caption{\textbf{\sys's fuzzing process example for existing vulnerability on Immunefi~\cite{immunefi_silo_2023}.}}
  \label{tab:code_operations}
\end{table*}

To evaluate the performance of \textsc{FAuditor}, we ran \textsc{FAuditor} alongside benchmark fuzzers on the performance dataset to compare their effectiveness in reproducing existing MEVuls (\textbf{RQ1}). Additionally, we present a case study that illustrates \textsc{FAuditor}'s detection pipeline and provides insights into its exploitation capabilities.

\begin{table}[ht]
\tiny
\rowcolors{3}{gray!10}{gray!30}
\resizebox{\linewidth}{!}{
\centering
\begin{tabular}{c c c c c c c c}
\hline
 & \multicolumn{4}{c}{\textbf{Type}} & \multicolumn{3}{c}{\textbf{Tool Results}} \\
\multirow{-2}{*}{\textbf{Project Name}} & \textbf{TA} & \textbf{BC} & \textbf{ES} & \textbf{VS} & \textbf{Smartian} & \textbf{Echidna} & \textbf{Ours} \\
\hline
DFX Finance & - & \checkmark & - & - & \textbf{TP} & FP & \textbf{TP} \\
Omni Protocol & - & \checkmark & - & - & \textbf{TP} & \textbf{TP} & \textbf{TP} \\
RocketPool & - & - & \checkmark & - & FP & T/O & \textbf{TP} \\
Lido & - & - & \checkmark & - & T/O & \textbf{TP} & \textbf{TP} \\
Cream Finance & - & \checkmark & - & - & \textbf{TP} & \textbf{TP} & \textbf{TP} \\
Harvest Finance & - & - & \checkmark & \checkmark & FP & T/O & \textbf{TP} \\
Uranium Finance & - & - & \checkmark & \checkmark & T/O & T/O & \textbf{TP} \\
BCoin & \checkmark & - & - & - & FP & FP & T/O \\
Alpha Homora & \checkmark & \checkmark & - & - & T/O & T/O & \textbf{TP} \\
\dots 129 more & \dots & \dots & \dots & \dots & \dots & \dots & \dots \\
\hline
Overall F-score &   &   &   &   & 0.29 & 0.52 & \textbf{0.94} \\
\hline
\end{tabular}
}
\caption{\textbf{Representative Detection Results of Existing MEVuls.}}
\label{tab:existing_big_contracts_bugs}
\end{table}
\PP{Results}
\autoref{tab:existing_big_contracts_bugs} presents the comparative results between \sys and benchmark fuzzers. 
The results demonstrate that \textsc{FAuditor} outperforms the others with an overall F-score of 0.94, whereas the best-performing alternative achieved only 0.52. 
\textsc{FAuditor} missed just 13 out of 138 MEVuls, while other benchmark fuzzers missed more than 82. 
This indicates that \textsc{FAuditor} can reliably reproduce existing vulnerabilities by precisely identifying their root causes, yielding a significantly higher F-score with better detection of MEVuls and a lower false positive rate, even after the compromise of true positive criteria for benchmark fuzzers. 

\PP{Case Study of FAuditor Reproducing MEVul}\label{sec:existing_study_case}
To further demonstrate \sys's concrete performance in practical scenarios, we showcase its automated procedure to reproduce a MEVul in the Silo lending protocol~\cite{silo_contract}, highlighting the tools' correctness and effectiveness in real-world. 
Its original exploitation credits to the security platform Immunefi~\cite{immunefi_silo_2023}. 

The Silo protocol~\cite{silo_contract} allows users to deposit collateral tokens and borrow other tokens. The deposit function mints collateral shares to the depositor and transfers tokens to the contract, updating the total amount in \texttt{\_assetStorage[\_asset]}. The borrowing function updates the interest rate, checks liquidity, and validates the Loan-to-Value (LTV) ratio to ensure the collateral's value is sufficient for the borrowed amount. The LTV is calculated as:
\begin{equation}
    \begin{aligned}
        &\texttt{user\_borrowed\_value}= \\
        &\frac{\texttt{user\_borrowed\_amount} \cdot\texttt{token1\_borrowed\_amount}}{\texttt{token1\_collateral\_amount}},\\
        &\texttt{user\_collateral\_value}=  \\
        &\frac{\texttt{user\_collateral\_amount} \cdot\texttt{token2\_borrowed\_amount}}{\texttt{token2\_collateral\_amount}}.\\
    \end{aligned}
\end{equation}
Generally, more collateral allows more borrowing, with the token price influenced by supply and demand dynamics. However, a fuzzer-identified vulnerability shows that improper token transfers can disrupt this balance.


\textsc{FAuditor} generated transactions (\autoref{tab:code_operations}) uncover a critical vulnerability where the contract's total value plummets due to an anomaly between transactions 6 and 8. The attack is initiated by an unexpected token transfer at transaction 3, exploiting the borrowing function's reliance on \texttt{ERC20().balanceOf()}. This creates an inflated view of the token's internal price, allowing the attacker to borrow more than the actual collateral value. As a result, when the contract owner attempts liquidation at transaction 9, the contract becomes insolvent due to the earlier exploitation.

What sets \textsc{FAuditor} apart from benchmark fuzzers is its precision in pinpointing exactly when and where value discrepancies occur, providing developers with detailed, actionable insights. In contrast, traditional fuzzers often produce vague alerts like \textit{Unexpected Ether} without offering sufficient context, leaving developers to dismiss them as false positives or inconsequential. This lack of specificity can cause developers to overlook critical vulnerabilities, which, as demonstrated here, can lead to severe financial losses.

\subsection{Discovering Zero-day MEVuls}\label{sec:exp_RQ2_novel}

To evaluate the ability of \sys to discover zero-day MEVuls, we ran \sys on 23 Dapps and 15,132 individual contracts.

\PP{Results}
We found 220 zero-day MEVuls, while even the most distinct benchmark fuzzers could only find 40.
We report all the zero-day MEVuls to the respective repository owners or the Etherscan platform.
As of the submission date, 12 of these vulnerabilities have been confirmed by the authors, and we remain in contact for subsequent mitigation efforts.
We show the detailed results and case study of zero-day MEVuls as below.

\noindent \textit{In Dapps:} From 23 Dapps, we found 7 contracts with MEVuls listed in the \autoref{tab:big_contracts_bugs}. We also tried benchmark fuzzers on these zero-day vulnerabilities, but none of them were able to detect the issues. They either produced irrelevant and non-exploitable results or timed out without any output.

\noindent \textit{In Individual Smart Contracts:} We found 213 money-related vulnerabilities from the 15,132 real-life individual smart contracts, as shown in the statistics results \autoref{tab:small_contracts_bugs}. The best-performed benchmark tool Smartian~\cite{Choi2021} can only report 40 (19\%) of these zero-day MEVuls.


\noindent \textbf{Mitigation.}
All the vulnerabilities mentioned above have been 100\% tested and triggered on our locally configured testnet. They have no impact on the mainnet and certainly do not cause any damage. As a mitigation measure, we have written the proof of concepts (PoCs) and attempted to contact the contract developers. Some of the developers have responded quickly and fixed the issues. Regarding the contracts that we couldn't find the author, we have reported the vulnerabilities to the Etherscan~\cite{noauthor_ethereum_nodate} platform.

\begin{table}[ht]
\tiny
\rowcolors{3}{gray!10}{gray!30}
\resizebox{\linewidth}{!}{
\centering
\begin{tabular}{c c c c c c c c}
\hline
 & \multicolumn{4}{c}{\textbf{Type}} & \multicolumn{3}{c}{\textbf{Tool Results}} \\
\multirow{-2}{*}{\textbf{Anonym. address}} & \textbf{TA} & \textbf{BC} & \textbf{ES} & \textbf{VS} & \textbf{Smartian} & \textbf{Echidna} & \textbf{Ours} \\
\hline
0x96cc...2fc8 & \checkmark & \checkmark & - & - & T/O & T/O & TP \\
0x9cc8...b852 & - & \checkmark & - & - & FP & FP & TP \\
0x18b7...7f38 & - & - & \checkmark & - & T/O & T/O & TP \\
0x7b74...9020 & - & - & \checkmark & \checkmark & T/O & FP & TP \\
0x6f80...b1f8 & - & - & \checkmark & \checkmark & FP & T/O & TP \\
0xc5d1...793d & - & \checkmark & - & \checkmark & T/O & FP & TP \\
0xd9fc...e07b & \checkmark & \checkmark & - & - & FP & FP & TP \\
\hline
Feasibility &   &   &   &   & \texttimes & \texttimes & \checkmark \\
\hline
\end{tabular}
}
\caption{\textbf{Detection Results of Zero-day MEVuls In Dapps.}}
\label{tab:big_contracts_bugs}
\end{table}

\begin{table}[ht]
\scriptsize
\rowcolors{2}{gray!10}{gray!30}
\resizebox{\linewidth}{!}{
  \centering
  \begin{tabular}{c c c c c c}
  \hline
  \textbf{Type}& \textbf{Total (Ours)} & \textbf{Smartian} & \textbf{ILF} & \textbf{Echidna} & \textbf{ItyFuzz} \\
  \hline
  \textbf{TA} & 64 & 25 & 12 & 2 & 19\\
  \textbf{BC} & 120 & 13 & 9 & 8 & 2\\
  \textbf{ES} & 17 & 1 & 4 & 0 & 4\\
  \textbf{VS} & 12 & 1 & 0 & 2 & 3\\
  \hline
  \textbf{Total} & \textbf{213} & 40 (19\%) & 25 (12\%) & 12 (6\%)& 28 (13\%)\\
  \hline
  \end{tabular}
}
  \caption{\textbf{Zero-day MEVuls in Individual Smart Contracts}}
  \label{tab:small_contracts_bugs}
\end{table}

\PP{Case Study of Zero-day Vulnerabilities in Dapps}
We present four representative examples of zero-day MEVuls, each demonstrating distinct vulnerability types, severe financial impact, and varied triggering mechanisms. These examples highlight the diversity in both the nature of the exploits and the methods required to uncover them.


\noindent \texttt{0x96ccXXX:} 
This lending protocol contract is vulnerable in a payable function that fails to update the total deposit amount and the user's share correctly when tokens are deposited. \textsc{FAuditor} detected this by checking Transfer Arrival and Balance Conservation violations. While only affecting its internal tokens other than transferable tokens, this vulnerability is considered medium severity as it doesn't cause further damage beyond token loss.

\noindent \texttt{0x9cc8XXX:} 
The contract's mint function, not guarded by \texttt{onlyOwner} but by \texttt{require} checking \texttt{tx.origin}, allows an attacker to mint tokens infinitely by detouring the transaction through their contract. \textsc{FAuditor} detected this through the Fund Value Stability violation. This is a critical vulnerability due to the potential for unlimited token creation.

\noindent \texttt{0x7b74XXX:} 
This DeFi protocol allows token deposits for borrowing or interest accrual. \textsc{FAuditor} detected vulnerabilities by directly transferring tokens, affecting Swap Rate Stability and Fund Value Stability, potentially causing unrestricted borrowing and project insolvency. This is a high-severity issue.

\noindent \texttt{0xc5d1XXX:} 
An integer overflow vulnerability in this contract's payable function leads to the user's share diminishing to zero after multiple large token transfers. This causes Transfer Arrival, Balance Conservation, and Fund Value Stability violations, rated as medium severity. Traditional tools might detect the overflow but not the financial impact.

\begin{figure*}[ht]
    \begin{subfigure}[b]{0.48\textwidth}
      \includegraphics[width=\textwidth]{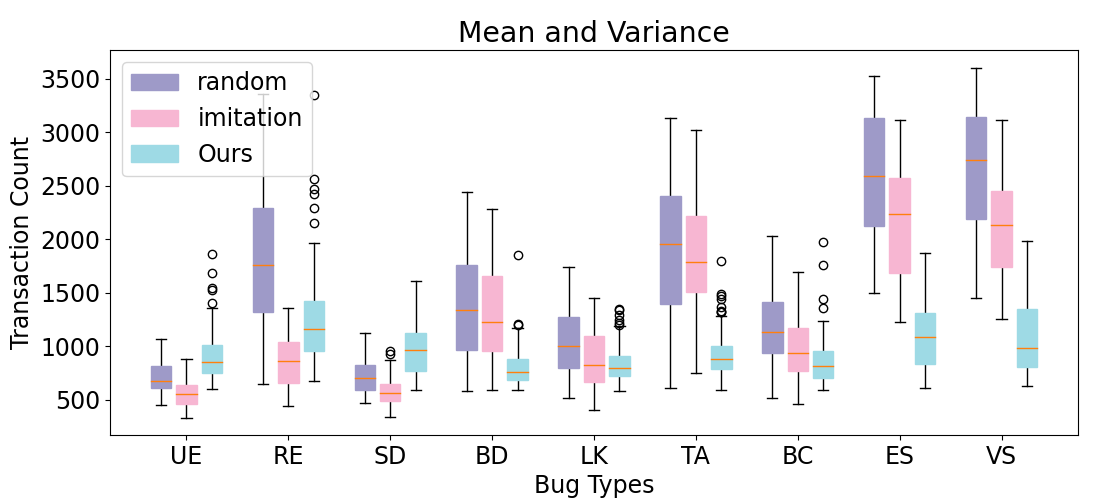}
      \caption{The Time Taken to Detect Vulnerabilities}
      \label{fig:bug_time}
    \end{subfigure}
    \begin{subfigure}[b]{0.48\textwidth}
      \includegraphics[width=\textwidth]{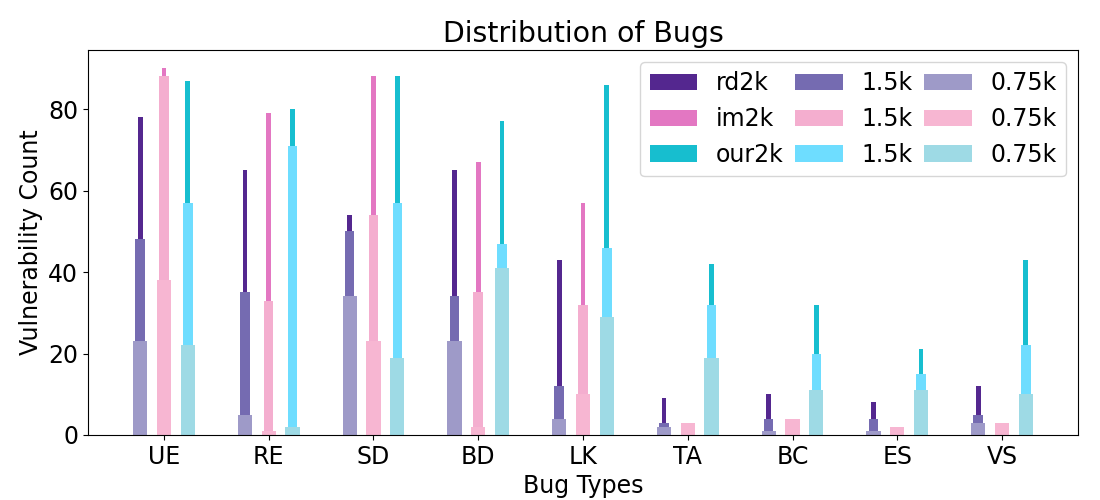}
      \caption{The Number of Vulnerabilities detected in Fixed Time}
      \label{fig:bug_count}
    \end{subfigure}
  \caption{\textbf{Vulnerability detection speed measured respectively with a fixed number of vulnerabilities or fixed amounts of time.}}
  \label{fig:bug_performance}
  \end{figure*}

\subsection{Code Coverage and Computational Overhead}
\label{sec:exp_RQ4_coverage}


The dataset for coverage and overhead evaluation includes over 15K real-life smart contracts. We assess \textsc{FAuditor}'s performance by comparing its instruction coverage to benchmark fuzzers. Instruction coverage offers finer granularity than basic block coverage, revealing more detailed execution paths, and both metrics yield consistent results during our testing. We prioritize instruction coverage to emphasize \textsc{FAuditor}'s depth of analysis.

\PP{Results}
The results in the \autoref{tab:coverage} show that \textsc{FAuditor} outperforms benchmark fuzzers regarding the final instruction coverage rate. \textsc{FAuditor}'s advantage on Dapps is remarkably 35\% higher than the state-of-the-art AI-driven tool ILF~\cite{he_learning_2019} which performs best on individual contracts shown in \autoref{tab:existing_big_contracts_bugs}. Regarding the computational overhead compared to ILF, it's 1.18\texttimes \ better in CPU and 1.34\texttimes \ better in memory consumption. 

The better coverage is achieved at the expense of overhead, which is still acceptable across all the benchmarks. The significant advantage of \textsc{FAuditor} on Dapps is attributed to our techniques of dynamically adapting to multiple EVM versions and rediscovering proxy-deployed contracts. The EVM Version Compatibility technique guarantees the tool can correctly register all the dependencies for the DApp, and the Contract Rediscovery technique ensures that the proxy-deployed contract is interactable while testing. Thus, more contracts and instructions are visible to \textsc{FAuditor} than in other benchmarks' environments.

\begin{table}[ht]
\rowcolors{3}{gray!10}{gray!30}
\scriptsize
\resizebox{\linewidth}{!}{
    \centering
    \begin{tabular}{lcccc}
        \toprule
        \multirow{2}{*}{\textbf{Tool}} & \multicolumn{2}{c}{\textbf{Final Instruction Coverage (\%)}} & \multicolumn{2}{c}{\textbf{Average Overhead}} \\
         & \textbf{Individual Contract} & \textbf{Dapp} & \textbf{CPU (\%)} & \textbf{Memory (kMb)}\\
        \midrule
        \textbf{Smartian} & 74.4 & 20.3 & 113 & \textbf{0.4}\\
        \textbf{ILF} & 90.9 & 62.9 & \textit{807} & \textit{5.9}\\
        \textbf{Echidna} & \textit{64.3} & 29.7 & \textbf{104} & 0.8 \\
        \textbf{IttyFuzz} & 91.7 & \textit{13.9} & 121 & 3.2\\
        \textbf{Ours} & \textbf{92.5} & \textbf{87.9} & 680 & 4.4\\
        \bottomrule
    \end{tabular}
}
    \caption{\textbf{Instruction Coverage and Overhead of \sys.}}
    \label{tab:coverage}
\end{table}


\subsection{Impact of the New Strategy on Detection Efficiency}\label{sec:exp_RQ3_speed}
To implement the evaluation, we instrumented benchmarks' oracles to align with ours while reserving their features of function generation strategy. We selected ILF~\cite{he_smart_2020}, a state-of-the-art AI-driven fuzzer known for its speed in detecting vulnerabilities, and included Echidna~\cite{Groce2021a}, the most widely-used fuzzer, which offers broad coverage and a straightforward random strategy, making it an ideal comparison point.

\PP{Results}
The dataset for performance evaluation comprises 556 labeled individual contracts. The vulnerability types include Unexpected Exception (UE), Re-entrance (RE), Suicidal (SD), Block State Dependency (BD), Leaking (LK), addition to our defined MEVul types Transfer Arrival (TA), Balance Conservation (BC), Exchange Rate Stable (ES), and Total Value Stable (VS). \autoref{fig:bug_time} indicates the average time to detect each type of vulnerability, e.g., for Echidna's random strategy, the average time to detect UE is about 700 transactions. In this result, \textsc{FAuditor} (blue bar) spends fewer transactions to detect vulnerabilities for most types. \autoref{fig:bug_count} indicates the total vulnerability number detected within a specific time range. For example, for Echidna's random ("rd" in the figure) strategy, about 22 UE were detected respectively, and each can be detected under 0.75k transactions. In this result, \textsc{FAuditor} detects more vulnerabilities in limited transaction numbers than ILF's and Echidna's strategies.

For the average time to detect vulnerabilities (\autoref{fig:bug_time}) and the number of vulnerabilities detected within a fixed time (\autoref{fig:bug_count}), \sys consistently outperforms ILF~\cite{he_smart_2020} and Echidna~\cite{Groce2021a} for the most of the vulnerability types, meeting expectations for MEVuls and even surpassing benchmarks in some traditional vulnerabilities.

\section{Related Work}
\label{sec:related_work}

\PP{Traditional Fuzzers}
Heuristic-based fuzzers~\cite{ityfuzz,wang2024sfuzz2,jiang_contractfuzzer_2018, Grieco2020, Groce2021a, fleischer2023actor} use heuristic methods to generate and prioritize test cases, laying the groundwork for smart contract fuzzing by ensuring broad input coverage. Symbolic execution-based fuzzers~\cite{torres_confuzzius_2021, Choi2021, noauthor_enzymefinanceoyente_nodate, noauthor_modules_nodate} systematically explore contract state spaces to find vulnerabilities. ConFuzzius~\cite{torres_confuzzius_2021} enhances fuzzing by combining symbolic execution with traditional fuzzing to improve code coverage, while Smartian~\cite{Choi2021} further optimizes this with path-targeted input generation.

\PP{AI-Enhanced Fuzzers}
Machine learning-driven fuzzers~\cite{he_learning_2019,su_effectively_2022,liu_rethinking_2023, bai2023guiding} have shown significant improvement in vulnerability detection. ILF~\cite{he_learning_2019} leverages machine learning to guide fuzzing, boosting vulnerability discovery rates, and RLF~\cite{su_effectively_2022} uses reinforcement learning to target vulnerable code dynamically. Recent LLM-based fuzzing approaches like TrustLLM~\cite{ma_combining_2024} and LLM4Fuzz~\cite{shou2024llm4fuzz} aim to generate input sequences for uncovering vulnerabilities, but they lack accessible implementations and validated results, limiting their comparison with our tool.

\section{Conclusion}
\label{sec:conclusion}

The paper introduces \textsc{FAuditor}, an advanced fuzzer designed for Monetarily Exploitable Vulnerabilities (MEVuls) in smart contracts by leveraging auditor knowledge and machine-learning techniques. It effectively simulates sophisticated attacks and uncovers new types of MEVuls.
The evaluation results confirm that \textsc{FAuditor} efficiently identifies known vulnerabilities and discovers new types of vulnerabilities previously undetected by state-of-the-art tools in real-world scenarios, especially in DApp environments. This represents a significant advancement in smart contract security, setting a new standard for automated smart contract auditing.





\bibliographystyle{IEEEtran}

\footnotesize
\bibliography{mybib}


\end{document}